\documentclass[10pt]{article}
\usepackage{amsmath,amssymb,latexsym,multirow,color,titlesec}
\usepackage[nosort]{cite}
\usepackage{mathtools}
\usepackage{graphicx, fancyhdr}    
\usepackage{amsfonts}              
\usepackage{subcaption}
\usepackage{amsthm}                
\usepackage[font=small,labelfont=bf]{caption}
\usepackage{booktabs, amscd}
\usepackage{float,graphicx}
\usepackage{rotating,longtable}
\usepackage{pdflscape}
\restylefloat{table}

\titleformat{\paragraph}
{\normalfont\normalsize\bfseries}{\theparagraph}{1em}{}
\titlespacing*{\paragraph}
{0pt}{3.25ex plus 1ex minus .2ex}{1.5ex plus .2ex}

\begin{document}

\title{Blind prediction of protein B-factor and flexibility}

\author{David Bramer$^{1}$ and
Guo-Wei Wei$^{1,2,3}$ \footnote{ Address correspondences  to Guo-Wei Wei. E-mail:wei@math.msu.edu}\\
$^1$ Department of Mathematics \\
Michigan State University, MI 48824, USA\\
$^2$  Department of Biochemistry and Molecular Biology\\
Michigan State University, MI 48824, USA \\
$^3$ Department of Electrical and Computer Engineering \\
Michigan State University, MI 48824, USA \\
}

\maketitle
\newcommand{\rd}[1]{{{#1}}}
\begin{abstract}
 
Debye-Waller factor, a measure of X-ray attenuation, can be experimentally observed in protein X-ray
crystallography. Previous theoretical models have made strong inroads in the analysis of B-factors by
linearly fitting protein B-factors from experimental data. However, the blind prediction of B-factors
for unknown proteins is an unsolved problem. This work integrates machine learning and advanced
graph theory, namely, multiscale weighted colored graphs (MWCGs), to blindly predict B-factors of
unknown proteins. MWCGs are local features that measure the intrinsic flexibility due to a protein structure. 
Global features that connect the B-factors of different proteins, e.g., the resolution of X-ray crystallography, 
are introduced to enable the cross-protein B-factor predictions. Several machine learning approaches, including 
ensemble methods and deep learning, are considered in the present work. The proposed method is validated with 
hundreds of thousands of experimental B-factors. Extensive numerical results indicate that the blind B-factor 
predictions obtained from the present method are more accurate than the least squares fittings using traditional
methods.

\textit{Keywords:} Weighted colored graph,     Protein flexibility,   Gradient boosted trees,     Random forest, Convolutional neural network.
\end{abstract}
\newpage


\section{Introduction}

Protein beta factor (B-factor) or temperature factor (Debye-Waller factor) is a measure of atomic   mean squared displacement or  uncertainty  in the X-ray scattering or  neutron scattering structure determination. For a given protein at a given temperature, a large B-factor is caused by the atomic thermal fluctuation and  low attenuation rate. The latter depends also on the experimental modality. For example, the hydrogen atom has a low  attenuation rate in X-ray scattering because of its small number of electrons but has a normal  attenuation rate for  neutron scattering.  For a given element type under the same experimental condition, the B-factor of an atom is determined by its intrinsic flexibility and possible crystal packing effects.   It has been previously shown that intrinsic   flexibility correlates to important protein conformational variations \cite{JMa:2005}. That is, protein structural fluctuation provides an important link between structure, and function of a protein. As such, accurate prediction of protein B-factors is an important and meaningful metric in understanding protein structure, flexibility and function  \cite{Frauenfelder:1991}.

One successful class of methods in protein B-factor prediction was those that used elastic mass-and-spring networks derived from Hooke's Law. These models represent the alpha carbons of biological macromolecules as a mass and spring network to predict B-factors based on a harmonic potential. Each alpha carbon in a protein is regarded as a node in the network, and edges are weighted based on a potential function. In these models, a pair of  nodes is connected by an edge if they fall within a predefined Euclidean cutoff distance. This approach captures the local  non-covalent interactions between an individual alpha carbon atom and nearby alpha carbon atoms.

Normal mode analysis (NMA) was one of the first mass-and-spring methods used for protein B-factor prediction. This method is independent of time and makes use of a Hamiltonian  matrix for atomic interactions. Here the modes of the system correspond to motion where all parts of the molecule are moving sinusoidally with the same frequency and phase. Moreover, eigenvalues of the system correspond to characteristic frequencies that correlate with protein B-factors. Low-frequency modes correlate with operative motions which can be useful for hinge detection. NMA has also been found to be useful in characterizing coarse grain deformation of supramolecular complexes. \cite{JMa:2005,Tasumi:1982,Brooks:1983,Levitt:1985}

The elastic network model (ENM) was introduced to reduce the computational cost of NMA by using a simplified spring network \cite{Tirion:1996}. One successful ENM model is the anisotropic network model (ANM). This model uses a simplified spring potential between each residue, then determines the modes of the system via matrix diagonalization. ANM still retains many of the insightful features of NMA but with a much lower computation cost.\cite{Atilgan:2001,Bahar:1997,Bahar:1998}

The Gaussian network model (GNM) was introduced as a simplified method for B-factor prediction \cite{Bahar:1997}. Similar to previous models, a graph network is constructed using alpha carbon as nodes and edges based on a prescribed cutoff distance. GNM uses a distance-based  Kirchhoff (or connectivity) matrix to represent the interaction between each two alpha carbon atoms (nodes). The expectation values of residue fluctuations or mean-square fluctuations are found in the diagonal terms of a covariance  matrix. GNM provides good-coarse grained results with relatively low computational cost. \cite{Haliloglu:1997}

More recently, the flexibility and rigidity index (FRI) methods have provided improved results. These methods construct graph centrality based on  radial basis functions which scale distance non-linearly \cite{KLXia:2013f}. Fast FRI (fFRI) provides a version of FRI with a very low computation cost while still maintaining satisfactory results  \cite{Opron:2014}.   Anisotropic FRI (aFRI) offers a matrix version of FRI to compute protein anisotropic motions.    Moreover, the multiscale flexibility rigidity index (mFRI) is able to capture protein multiscale interactions using several radial basis functions with different parameterizations \cite{Opron:2015a,DDNguyen:2016b}.

Previously the authors introduced a multiscale weighted colored graph (MWCG) model for protein flexibility analysis \cite{Bramer:2018}.  The MWCG is a geometric graph model that offers the most accurate and reliable protein flexibility analysis and B-factor prediction to date.  It is about 40\% more accurate than GNM \cite{Bramer:2018}. The basic idea of MWCG is to color (label) a protein graph based on element interaction types. Each atom of given an element type selection represents a graph vertex and subgraphs are defined according to specific heavy element types. A generalized centrality is defined for each subgraph vertex. Using various parameterizations of radial basis functions, this method is able to capture  multiscale   element specific interactions. The MWCG method can be combined with various earlier FRI approaches, such as fFRI, mFRI and aFRI, to further strengthen its power in the analysis of intrinsic protein flexibility. Additionally, MWCG works well not only for C$_\alpha$ carbons but also for all the atoms in a protein, i.e., non-C$_\alpha$ carbon, nitrogen, oxygen, and sulfur atoms. Hydrogen atoms can be treated similarly if they are available in the dataset \cite{Bramer:2018}. 

All of the aforementioned methods are designed for the analysis of intrinsic protein flexibility due to protein structure and packing crystal packing. However, none was designed to predict the B-factors of an unknown protein. Indeed, all of these methods fit experimental B-factors of given protein by the least squares algorithm. They generally do a poor job in predicting flexibility across proteins. Stated differently, the fitting coefficients  obtained from one protein are not applicable to a different protein in general.
 This is largely due to the fact that protein B-factor depends also on a large number of effects, including X-ray crystal quality, crystal  symmetry (i.e., space group), data collecting method,   data collecting  environment, equipment condition, etc. 
 Consequently, the blind prediction of  protein flexibility and B-factors remains a major challenge.   

Recently, advances in graphics processing unit (GPU) computing and optimization have led to impressive biophysical predictions for various problems using machine learning, particularly, deep learning techniques. In this work,  we propose  machine-learning based methods for blind protein B-factor predictions. We introduce two sets of features, the global ones, and local ones. Global features are designed to represent crystal and experimental conditions across different proteins, while local features are devoted to describing structural and atomic properties within a protein structure. We compile and engineer local and global features from a large set of known protein data as a training set, then apply machine learning  techniques to establish regression models which are used  for the blind prediction of B-factors of unknown protein structures. In terms of machine learning procedures, we use a variety of local and global protein features of a labeled training set  to construct regression models that can blindly predict the B-factors of a test set, consisting of entirely new proteins. In this work, we explore the random forest, boosted gradient decision trees, and deep learning methods  for blind protein B-factor predictions. Using a large and diverse set of proteins from the protein data bank ensures technical robustness. In addition to previously explored features such as MWCG kernels and element type, we also include secondary structural information and local packing density features to further improve our results.


\section{Methods and algorithms}\label{sect:Theory}
The success of blind protein B-factor predictions depends crucially on the representation of biomolecular structures. We employ  MWCGs as local features to describe protein structures. A brief review of MWCGs is given below. 
 
\subsection{Multiscale weighted colored graphs}\label{sect:MWCG}

Graph  theory concerns the relationship of a set of vertices, denoted as  $V$, in terms of pairwise connectivity, i.e., edges  $E$.  We use a graph to describe the non-covalent interactions in proteins.  To improve our graph theory representation,  we consider colored graphs in which different types of elements are labeled.   We classify labeled  protein atoms into subgraphs where colored edges correspond to element specific interactions. Specifically, we label the $i$th atom by its element type $\alpha_j$ and position ${\bf r}_j$. As such,  vertices are labeled as \[V = \{(\mathbf{r}_j, \alpha_j)|\mathbf{r}_j\in {\rm I\!R}^3; \alpha_j \in {\cal C};  j=1,2,\ldots,N \}\], where ${\cal C}=$\{C, N, O, S \} are the set of elements whose pairwise interactions will be considered. Hydrogen is omitted from this list due to its absence from most PDB data  and can be added without affecting the present description. The set of edges in the colored protein graph are  element specific pairs  ${\cal P}=$\{CC,CN,CO,CS,NC,NN,NO,NS,OC,ON,OO,OS,SC,SN,SO,SS\}. For example,  the subset ${\cal P}_3=$\{CO\} contains all directed  CO pairs in the protein such that the first atom is a carbon and the second one is a nitrogen.  The direction  is maintained because the edge, $E$, is a set of weighted and  directed  interaction kernels of various pairs of atoms,      
\begin{equation}
E=\{\Phi^k(||\mathbf{r}_i-\mathbf{r}_j||; \eta_{ij})| (\alpha_i \alpha_j) \in {\cal P}_k;  k=1,2,\ldots,10;  i,j = 1,2,\ldots,N \},
\end{equation} 
 where $||\mathbf{r}_i-\mathbf{r}_j||$ is the Euclidean distance between the $i^{th}$ and $j^{th}$ atoms, $\eta_{ij}$ is a characteristic distance between the atoms and  $(\alpha_i \alpha_j)$ is a  directed  pair of element types. 
 Here $\Phi^k$ is a correlation function and is chosen to have the following properties  \cite{Opron:2014}
\begin{equation}
\Phi^k(||\mathbf{r}_i-\mathbf{r}_j||;\eta_{ij})=1, \mbox{ as } ||\mathbf{r}_i-\mathbf{r}_j||\rightarrow 0
\quad \quad (\alpha_i \alpha_j)   \in {\cal P}_k,
\end{equation}

\begin{equation}
\Phi^k(||\mathbf{r}_i-\mathbf{r}_j||;\eta_{ij})=0 \mbox{ as } ||\mathbf{r}_i-\mathbf{r}_j||\rightarrow\infty, \quad \quad (\alpha_i \alpha_j)   \in {\cal P}_k.
\end{equation} 
Our previous work   \cite{Opron:2014} has shown that generalized exponential functions,
\begin{equation}
\Phi^k(||\mathbf{r}_i-\mathbf{r}_j||;\eta_{ij})= e^{-(||\mathbf{r}_i-\mathbf{r}_j||/\eta_{ij})^{\kappa}},\quad (\alpha_i \alpha_j) \in {\cal P}_k; \quad \kappa>0, 
\end{equation} 
and generalized Lorentz functions,
\begin{equation} \Phi^k(||\mathbf{r}_i-\mathbf{r}_j||;\eta_{ij})= \dfrac{1}{1+(||\mathbf{r}_i-\mathbf{r}_j||/\eta_{ij})^{\nu}},\quad (\alpha_i \alpha_j)\in {\cal P}_k;\quad \nu>0, 
\end{equation} 
are good choices which satisfy the   assumptions. 


The centrality metric used in this work is an extension of harmonic centrality to subgraphs with weighted edges defined by the generalized correlation functions
\begin{equation}\label{ESRI}
\mu^k_i=\sum_{j=1}^{N}w_{ij}\Phi^k(||\mathbf{r}_i-\mathbf{r}_j||;\eta_{ij}), 
\quad (\alpha_i \alpha_j) \in {\cal P}_k,
\quad \forall i=1,2,\ldots, N,
\end{equation} 
where $w_{ij}$ is a weight function related to the element type. The WCG centrality in Eq. (\ref{ESRI}) describes the atomic specific rigidity which measures the stiffness  at the $i^{th}$ atom due to the $k$th set of contact atoms.

To characterize protein multiscale interactions, we use the atomic specific rigidity index from multiscale weighted colored graphs (MWCGs) introduced in our previous work \cite{Bramer:2018}.  
The  atomic rigidity  of  $i^{th}$ atom  at $n^{th}$ scale due to the $k^{th}$ set of interaction atoms is defined as
\begin{equation} \label{subs:CNO} 
\mu^{k,n}_i={\sum_{j=1}^{N}w_{ij}^n\Phi^k (||\mathbf{r}_i-\mathbf{r}_j||;\eta_{ij}^n)},\quad (\alpha_i \alpha_j)   \in {\cal P}_k,
\end{equation} 
where  $\Phi^k (||\mathbf{r}_i-\mathbf{r}_j||;\eta_{ij}^n)$ is a correlation kernel,  $\eta_{ij}^n$ a scale parameter, and $w^n_{ij}$ is an atomic type dependent parameter. We set $w^n_{ij}=1$ in the present work. 

While sulfur atoms play an important role in proteins they are so sparse that their kernels have a negligible effect on the current model. 
  Therefore, it is convenient to consider a subset of ${\cal P}$ in practical computations
\begin{equation}
{\cal\hat{P}}=\mbox{\big\{CC, CN, CO, NC, NN, NO, OC, ON, OO
\big\}}.
\end{equation}
We chose only C, N, and O element types due to their high occurrence frequency and important biological relevance.

\subsection{Machine learning features}\label{sect:ML}

%
\subsubsection{Global features}\label{feat}

Protein Databank (PDB) \*.pdb files provide the spatial atomic coordinates and the B-factor of  each atom in a protein as well as a variety of other types of observed data that can be used as features. In addition to the use of PDB spatial coordinates, this work makes use of global features provided in PDB files such as R-value, resolution, and number of heavy atoms. R-value and resolution are global measures of the quality of the atomic model obtained from crystallographic data. Another global feature we consider is the overall protein size. To allow the models to distinguish proteins of different sizes we use one hot encoding with the 10 size ranges 
 
\[[500,750,1000,1500,2000,2500,3000,4000,5000,30000],\]
 
where a protein element feature size will take on 1 if the number of heavy atoms (carbon, nitrogen, or oxygen) in that protein is less than or equal to the corresponding size and zero for the remaining sizes. For example, a protein with 1700 heavy elements would have the feature size vector for all of its atoms  given by
\[  [0 , 0 ,0   ,0   ,1   ,0   ,0	,0	 ,0	  ,0],\].
 
\begin{figure}[H]
\centering
\includegraphics[width=0.65\textwidth]{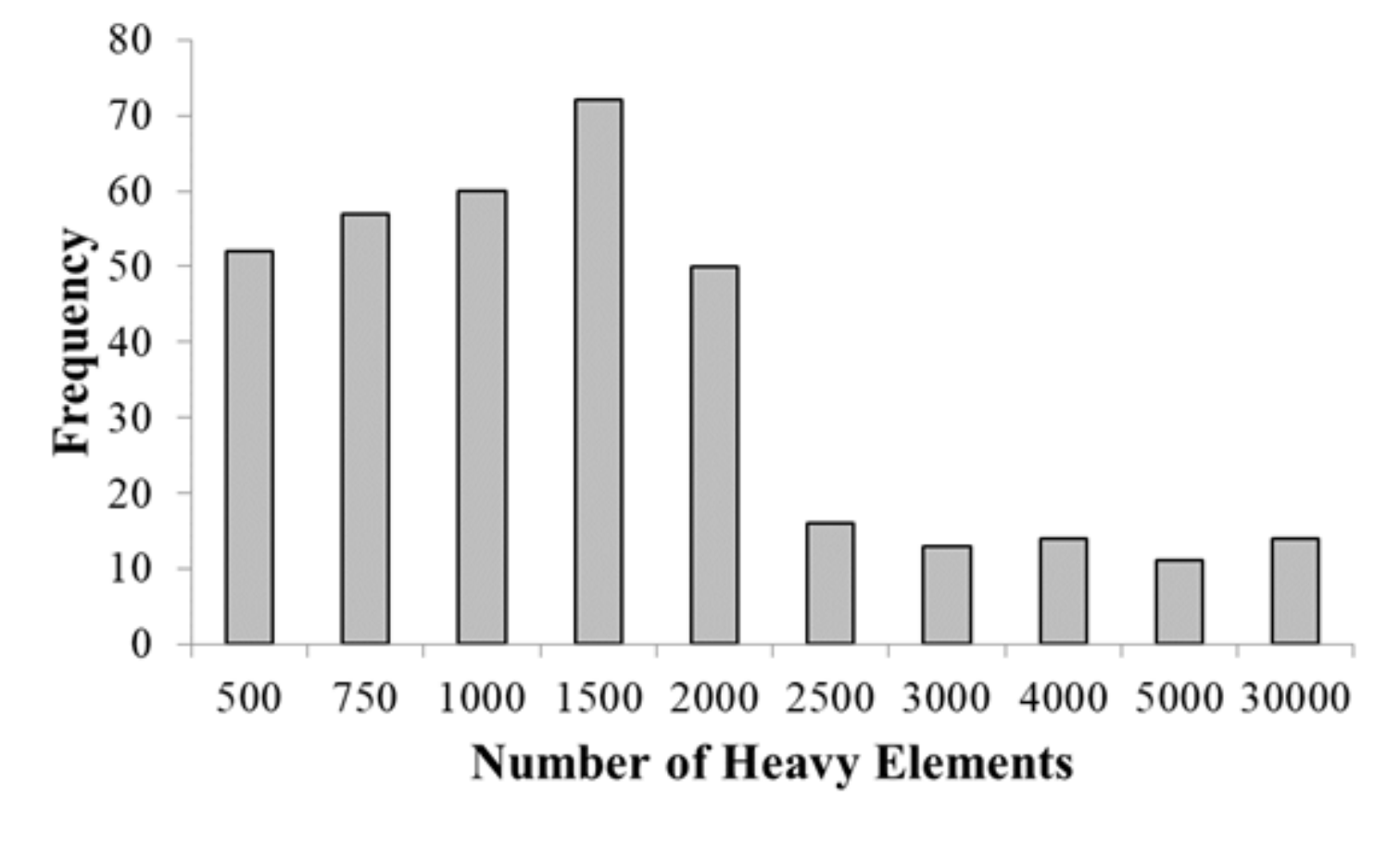}
\caption{Frequency of the number of heavy elements within  the proteins from the 364 protein dataset.}
\label{fig:freq}
\end{figure}

A frequency distribution of the size categories is provided in Figure \ref{fig:freq}. There are a total of 12 global protein size features.

\subsubsection{Local features}\label{feat_1}
PDB files also contain amino acid information for each element. Using one hot encoding we include amino acid information for each heavy element which results in 20 amino acid features. Similarly we one hot code the 4 different heavy element types carbon, nitrogen, oxygen, and sulfur for each element resulting in 4 additional features.

We use MWCG rigidity index described in Section \ref{sect:MWCG} to create feature vectors for carbon, nitrogen, and oxygen interactions with each element. Moreover, to capture multiscale interactions we use 3 different kernel choices for each interaction type. This results in a total of 9 MWCG feature vectors. The parametrization of the kernels is chosen based on our previous work and is provided in  Table \ref{tab:eta_op}.\cite{Bramer:2018}
\begin{table}[H]
\centering
\begin{tabular}{lccc}\hline\hline
Kernel Type 	    & $\kappa$& $\eta^n$ & $\nu$ \\ \hline 
Lorentz ($n=1$) 	& -	   &  16	   & 3\\
Lorentz ($n=2$)     & -    &  2        & 1 \\
Exponential ($n=3$) & 1    &  31      & -\\ \hline\hline
\end{tabular}
\caption{Parameters used for correlation kernels in a parameter-free MWCG based on previous results.\cite{Bramer:2018}}
\label{tab:eta_op}
\end{table}

The MWCG rigidity kernels do not entirely capture the density of nearby atoms. In this work, we define short, medium and long packing density features for each heavy atom. The packing density of the $i^{th}$ atom is defined as
\[p_i^d=\dfrac{N_d}{N},\]
where $d$ is the given cutoff in angstroms, $N_{d}$ is the number of atoms within the Euclidean distance of the cutoff to the $i^{th}$ atom, and $N$ the total number of heavy atoms of the protein. The packing density cutoffs used in this work are provided in Table \ref{tab:PD}.

\begin{table}[H]
\centering
\begin{tabular}{c|c|c}\hline\hline
 Short & Medium 	 & Long \\ \hline
 $d<3$ & $3\leq d<5$ &$5\leq d$ \\ \hline\hline
\end{tabular}
\caption{Packing density parameters in distance ($d$ \AA).}
\label{tab:PD}
\end{table}

We include secondary structural information generated using the STRIDE software. The STRIDE software provides secondary structural information about a protein given its atomic coordinates as a PDB file. STRIDE designates each atom as belonging to a helix (alpha helix,  3-10 helix, PI-helix), extended conformation, isolated bridge, turn, or coil. Additionally, STRIDE provides $\phi$ and $\psi$ angles and residue solvent accessible area \cite{Heinig2004}. Taken together this provides 12 secondary features.

\subsubsection{MWCG inputs}\label{feat_2}

Using the MWCG method, we apply Lorentz and exponential radial basis functions to construct multi-scale images for each element of a protein. To capture a large variety of scales we construct multiscale kernels for each heavy atom of a protein using various values of $\kappa$, $\nu$, and $\eta$. In particular we use $$\eta =\{1,2,3,4,5,10,15,20\}$$ and $$\kappa,\ \nu =\{ 2,2.5,3,3.5,4,4.5,5,5.5,6,6.5,7,8,9,10,11\}.$$

This results in 2D MWCG images of dimension $(8,30)$. We create images for all carbon, nitrogen, and oxygen interactions for each heavy atom giving each image three channels.

The image matrix is given by $F_i^k$ in equation \ref{mat:1}, where each atom $f_i^k(l,m,n)$ represents the flexibility index of the $i^{th}$ atom, and $k^{th}$ atom interaction (C, N, or O), $l=\eta$, $m=\{\kappa,\ \nu\}$, and $n$ the type of radial basis function. Values of $n=1$ and $n=2$ correspond to exponential and Lorentz radial basis functions respectively. 

\newcommand{\nm}[1]{f_i^k(#1)}
\newcommand\undermat[2]{%
  \makebox[0pt][l]{$\smash{\underbrace{\phantom{%
    \begin{matrix}#2\end{matrix}}}_{\text{$#1$}}}$}#2}
{    \small
\begin{equation}\left. F_i^k = \begin{bmatrix}
\nm{1,2,1} & \nm{1,2.5,1} & \ldots & \nm{1,11,1} & \nm{1,2,2}& \nm{1,2.5,2}& \ldots & \nm{1,11,2}\\
\nm{2,2,1} & \nm{2,2.5,1} & \ldots & \nm{2,11,1} & \nm{2,2,2}& \nm{2,2.5,2}& \ldots & \nm{2,11,2}\\
\vdots 	  &	 			&		 & 			  &			 &		  & \vdots \\
\nm{15,2,1} & \nm{15,2.5,1} & \ldots & \nm{15,11,1} & \nm{15,2,2}& \nm{15,2.5,2}& \ldots & \nm{15,11,2}\\
\undermat{\kappa}{\nm{20,2,1} & \nm{20,2.5,1} & \ldots & \nm{20,11,1} }& \undermat{\nu}{\nm{20,2,2}& \nm{20,2.5,2}& \ldots & \nm{20,11,2}}\\
\end{bmatrix}\right\}\eta
\label{mat:1}
\end{equation}
}

\subsection{Machine learning algorithms}

\rd{
A grid search was implemented for each method to determine the hyperparameters provided in Sections \ref{rf}, \ref{gbt} and \ref{deep}.
}
\subsubsection{Random forest}\label{rf}
Random forests are ensemble methods that can be used for either classification or regression tasks. Since protein B-factor is a continuous measurement, B-factor prediction is a regression task. Random forests  use a forest of $n$ decision trees, and in the regression task, the prediction output is the mean prediction of all the trees. Random forests have the added benefit of avoiding overfitting. Random forests are also invariant to scaling and can rank the importance of features used in the model.  Random forests are very robust to use for small- and medium-sized data sets. 

The number of $n$ trees used generally improves the predictive power of a random forest model but if $n$ is too large the model is susceptible overfitting the data set. In this work, we tested a variety of values for $n$ to find a balance between performance and cost.

\subsubsection{Gradient boosted trees}\label{gbt}

Gradient boosting is another ensemble method that assembles  a number of so called weak ``learners''   into a prediction model iteratively. Gradient boosting tree is a gradient descent method that optimizes  an arbitrary differentiable loss function to minimize 
the residuals from each step. Gradient boosted trees incorporate decision trees at each step of the gradient boosting to improve the predictive power of gradient boosting. Gradient boosted   trees are advantageous because they can handle heterogeneous features, have strong predictive power, and are generally robust to outliers.

The gradient boosted tree method has several hyper parameters that can be tuned. In this work we optimize the hyper parameters using the standard practice of a grid search. The parameters used for testing are provided in Table \ref{tab:param}. \rd{ Any hyper parameters not listed below were taken to be the default values provided by the python scikit-learn package.}

\begin{table}[H]
\centering
\caption{Boosted gradient tree parameters used for testing. Parameters were determined using a grid search. Any hyper parameters not listed below were taken to be the default values provided by the python scikit-learn package.}
\begin{tabular}{l|l} \hline \hline
Parameter 	   	  & Setting \\ \hline
Loss Function	  & Quantile\\ 
Alpha			  & 0.95\\ 
Estimators 	 	  & 1000\\ 
Learning Rate 	  & 0.001\\
Max Depth 		  & 4\\ 
Min Samples Leaf  & 9\\ 
Min Samples Split & 9\\ \hline\hline
\end{tabular}
\label{tab:param}
\end{table}

\subsubsection{Deep learning}\label{deep}

Neural networks initially intend to model the way neurons function in the brain. In a neural network, a batch of signals or feature inputs is passed through activation functions called perceptrons which are the functional units of the network. The weights of the networks are then trained using a loss function over several epochs. Each epoch passes the training data set through the network updating the weights according to the loss function. A neural network is considered deep when it has several ``hidden'' layers of perceptrons.

\begin{figure}
\centering
\includegraphics[scale=0.65]{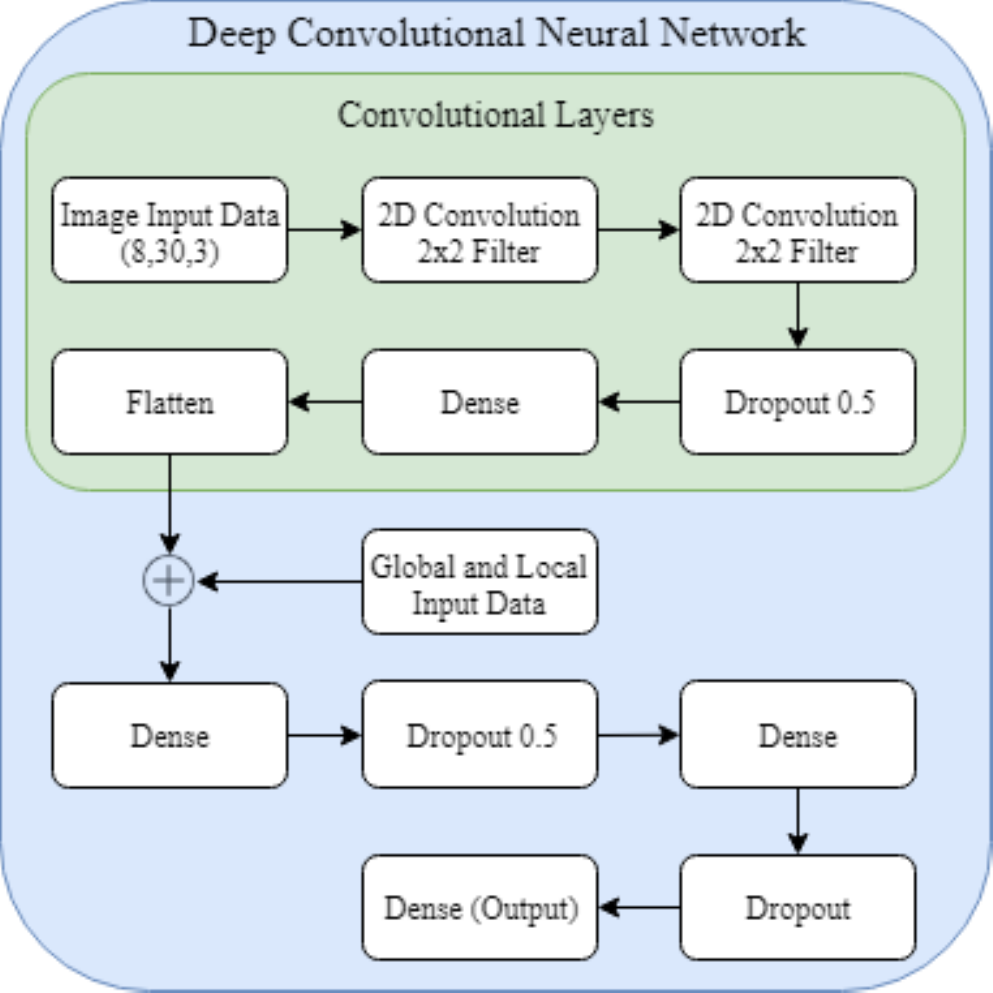}
\caption{The deep learning architecture using a convolutional neural network combined with a deep neural network. The plus symbol represents the concatenation of data sets. 
}
\label{fig:CNN}
\end{figure}

Convolutional neural networks (CNNs) have recently succeeded in classifying images. CNN's can extract features from images using convolutions with a pre-defined filter size. CNNs are advantageous because they can provide similar results without training the network on the full data set. In practice, one can extract high-level features by using several convolutions. In this work, we explore using a heat map of rigidity indices generated by three channel MWCG image features. We then merge the CNN output into a neural network that contains additional global and  local protein features. A diagram of the CNN architecture is given in Figure \ref{fig:CNN}.

The input of the CNN is a three-channel MWCG image of dimension (8,30,3). The model then applies two convolutional layers with 2x2 filters followed by a dropout layer at 0.5. This is followed by a dense layer which is flattened then joined with the other global and local features into a dense layer of 59 neurons followed by a dropout layer of 0.5, another dense layer of 100 neurons, a dropout layer of 0.25, a dense layer of 10 neurons, and finishes with a dense layer of 1 neuron. This results in a total of 21,584 trainable parameters for our network. 

\rd{The convolutional neural network (CNN)has several hyper parameters that can be tuned. In this work we optimize the hyper parameters using the standard practice of a grid search. The parameters used for testing are provided in Table \ref{tab:param_CNN}. Any hyper parameters not listed below were taken to be the default values provided by the python Keras package. }

\begin{table}[H]
\centering
\caption{Convolutional Neural Network (CNN) parameters used for testing. Parameters were determined using a grid search. Any hyper parameters not listed below were taken to be the default values provided by python with the Keras package.}
\begin{tabular}{ll} \hline\hline
Parameter 	   	  & Setting \\ \hline
Learning Rate 	  & 0.001\\
Epoch 	 	  	  & 100\\
Batch Size		  & 100\\
Loss			  & Mean Absolute Error \\ 
Optimizer		  & Adam \\ \hline\hline
\end{tabular}
\label{tab:param_CNN}
\end{table}

\subsubsection{Training set and test set}\label{TrainingParameters}
 \rd{The RF, GBT, and CNN were all trained and tested in the same manner. For each protein, a machine learning model is built using the entire dataset but excluding data from the protein whose B-factors are to be predicted. Overall, there are more than  620,000 atoms in our dataset. For each protein, this provides a training set of roughly 600,000 data points (i.e., atoms). For each heavy atom, there is a set of features as described in  Section \ref{sect:ML} and a B-factor value (label).  The features and the labels in the training set are used to train each machine learning model. Since we perform leave-one-out predictions, data from each protein is taken as a test set when its B-factors are to be blindly predicted.   } 

We implement random forest and boosted gradient models using the scikit-learn python package. For the CNN model, we also use the python package Keras with tensorflow as a backend.

\subsection{Datasets}\label{datasets}
Our study uses two datasets, one  from Park, Jernigan, and Wu  \cite{JKPark:2013} and the other from Refs. \cite{Opron:2014,Opron:2015a}.  The first  contains 3 subsets of small, medium, and large proteins  \cite{JKPark:2013} and the latter contains 364 proteins \cite{Opron:2014,Opron:2015a}.  The latter dataset is an extended version of the first. In these proteins, all sequences have a resolution of 3 {\AA} or higher and an average resolution of 1.3  {\AA}  and the sets include proteins that range from 4 to 3912 residues  \cite{JKPark:2013}.

For the CNN the feature datasets were standardized with mean 0 and variance of 1. Proteins 1OB4, 1OB7, 2OLX, and 3MD5 are excluded from the data set because the STRIDE software is unable to provide features for these proteins. We exclude protein 1AGN due to known problems with this protein data. Proteins 1NKO, 2OCT, and 3FVA are also excluded because these proteins have residues with B-factors reported as zero, which is unphysical.


\section{Results and discussions}
\label{sect:Num}

\subsection{Evaluation metric} \label{kernels}

\rd{
We successfully executed a leave-one-(protein)-out method to blindly predict the B-factors of all carbon, nitrogen, and oxygen atoms present in a given protein. For a comparison with other existing method,  we also list results for  predicted C$_\alpha$ B-factors, which are predicted in the same way as other heavy atoms. Machine learning was used to train a B-factor prediction model using the structural and B-factor data from a training data set as described in Sections \ref{TrainingParameters} and \ref{datasets}. The model was then used to predict the B-factors of all heavy atoms in a given protein using only its structural data.
}

To quantitatively assess our method for B-factor prediction we used the Pearson correlation coefficient given by

\begin{equation}
{\rm PCC}=\dfrac{\displaystyle\sum_{i=1}^{N}(B_i^e-\bar{B}^e)(B_i^t-\bar{B}^t)}{\bigg[\displaystyle\sum_{i=1}^{N}(B_i^e-\bar{B}^e)^2\displaystyle\sum_{i=1}^{N}(B_i^t-\bar{B}^t)^2\bigg]^{1/2}},
\end{equation}
where $B^t_i,i = 1, 2,\ldots,N$ 
are  predicted B-factors using the proposed method and $B^e_i ,i = 1, 2,\ldots,N$ experimental B-factors from the PDB file. The terms $B^t_i$ and $B^e_i$ represent the $i^{th}$ theoretical and experimental B-factors respectively. Here $\bar{B}^e$ and  $\bar{B}^t$ are averaged B-factors.  

\begin{figure}
\centering
\includegraphics[scale=0.75]{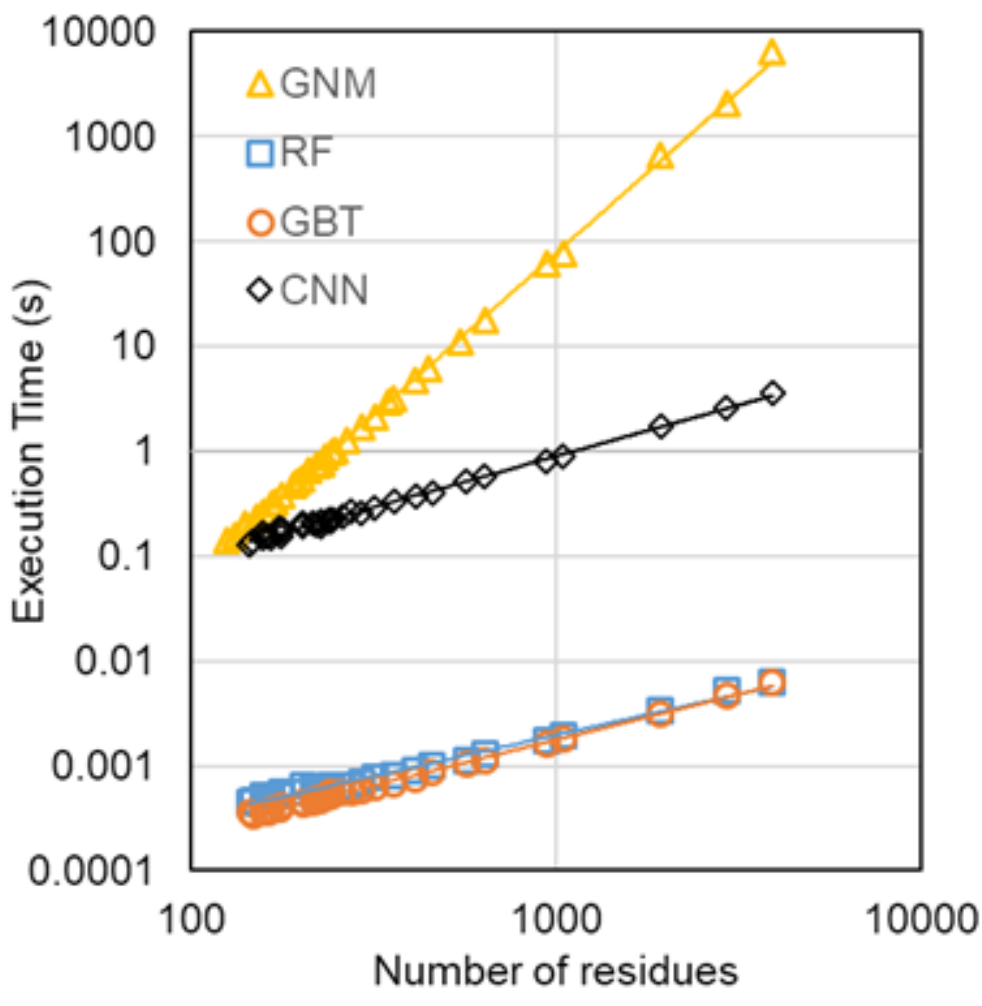}
\caption{CPU Efficiency comparison between GNM \cite{Opron:2014}, RF, GBT, and CNN algorithms. Execution times in seconds (s) versus number of residues. A set of 34 proteins, listed in Table \ref{tab:comp_time}, were used to evaluate the computational complexity.}
\label{fig:comp_time}
\end{figure}

\subsection{Efficiency comparison}
\rd{
 Computational efficiency in B-factor predictions is an important consideration for large proteins. Table \ref{tab:comp_time} lists the running times of GNM, RF, GBT, and CNN in our python implementations.  These results are depicted in  Figure \ref{fig:comp_time}. The proteins used to evaluate the computational complexity were the same as those used by Opron {\it et all} \cite{Opron:2014}. For this comparison we only predict B-factors for C$_{\alpha}$ atoms. Several proteins were excluded as GNM takes significantly too much CPU time to run. Tests excluded the time it took to load PDB files and feature data. The machine learning algorithm times exclude the training of the model, which, once trained, can be used for the prediction of all proteins. The results show that GNM has computational complexity of roughly $\mathcal{O}(N^3)$ due to the matrix decomposition while the ML algorithms are close to $\mathcal{O}(N)$, with $N$ being the number of atoms. The lines of best fit for CPU time ($t$) are $t\approx (4\times 10^{-8})* N^{3.09}$ for GNM, $t\approx (9\times 10^{-6})* N^{0.78}$ for RF, $t\approx (4\times 10^{-6})* N^{0.87}$ for GBT, and $t\approx(1.1\times 10^{-3})* N^{0.97}$ for CNN. 
}

\pagebreak

{\small
\begin{table}[H]
\centering
\begin{tabular}{cccccc}
\caption{CPU execution times, in seconds, from efficiency comparison between GNM \cite{Opron:2014}, RF, GBT, and CNN.}
\label{tab:comp_time} \\
\hline\hline \\
PDB	&	N	&	GNM\cite{Opron:2014}	&	RF	&	GBT	&	CNN	\\ \hline
3P6J	&	125	&	0.141	&	0.000455	&	0.000358	&	0.130	\\ 
3R87	&	132	&	0.156	&	0.000464	&	0.000339	&	0.138	\\ 
3KBE	&	140	&	0.187	&	0.000505	&	0.000384	&	0.149	\\ 
1TZV	&	141	&	0.203	&	0.000473	&	0.000365	&	0.163	\\ 
2VY8	&	149	&	0.219	&	0.000486	&	0.000359	&	0.156	\\ 
3ZIT	&	152	&	0.234	&	0.000519	&	0.000365	&	0.148	\\ 
2FG1	&	157	&	0.265	&	0.000518	&	0.000403	&	0.174	\\ 
2X3M	&	166	&	0.312	&	0.000526	&	0.000382	&	0.182	\\ 
3LAA	&	169	&	0.327	&	0.000514	&	0.000405	&	0.155	\\ 
3M8J	&	178	&	0.375	&	0.000548	&	0.000412	&	0.178	\\ 
2GZQ	&	191	&	0.468	&	0.000647	&	0.000454	&	0.195	\\ 
4G7X	&	194	&	0.499	&	0.000631	&	0.000445	&	0.209	\\ 
2J9W	&	200	&	0.546	&	0.000554	&	0.000424	&	0.208	\\ 
3TUA	&	210	&	0.655	&	0.000602	&	0.000472	&	0.217	\\ 
1U9C	&	221	&	0.733	&	0.000592	&	0.000486	&	0.198	\\ 
3ZRX	&	221	&	0.718	&	0.000654	&	0.000515	&	0.216	\\ 
3K6Y	&	227	&	0.765	&	0.000619	&	0.000490	&	0.189	\\ 
3OQY	&	234	&	0.873	&	0.000619	&	0.000502	&	0.211	\\ 
2J32	&	244	&	0.967	&	0.000625	&	0.000556	&	0.225	\\ 
3M3P	&	249	&	1.029	&	0.000621	&	0.000525	&	0.220	\\ 
1U7I	&	267	&	1.263	&	0.000647	&	0.000551	&	0.237	\\ 
4B9G	&	292	&	1.669	&	0.000693	&	0.000574	&	0.256	\\ 
4ERY	&	318	&	2.122	&	0.000775	&	0.000619	&	0.289	\\ 
3MGN	&	348	&	2.902	&	0.000655	&	0.000552	&	0.267	\\ 
2ZU1	&	360	&	3.136	&	0.000816	&	0.000675	&	0.337	\\ 
2Q52	&	412	&	4.696	&	0.000900	&	0.000750	&	0.369	\\ 
4F01	&	448	&	6.178	&	0.001016	&	0.000878	&	0.401	\\ 
3DRF	&	547	&	11.154	&	0.001131	&	0.001033	&	0.512	\\ 
3UR8	&	637	&	17.409	&	0.001307	&	0.001136	&	0.583	\\ 
2AH1	&	939	&	61.012	&	0.001716	&	0.001605	&	0.800	\\ 
1GCO	&	1044	&	75.801	&	0.001936	&	0.001814	&	0.905	\\ 
1F8R	&	1932	&	654.127	&	0.003343	&	0.003163	&	1.745	\\ 
1H6V	&	2927	&	2085.842	&	0.005205	&	0.004739	&	2.543	\\ 
1QKI	&	3912	&	6365.668	&	0.006261	&	0.006198	&	3.560	\\ \hline\hline
\end{tabular}
\end{table}
}

\subsection{Machine learning performance}


\rd{The results in Table \ref{tab:Avg_CC} show that for the blind prediction of all heavy atoms the convolutional neural network method performs best with an overall average Pearson correlation coefficient of 0.69. The gradient boosted and random forest ensemble methods performed similarly with Pearson correlation coefficients of 0.63 and 0.59 respectively.   For comparison, Table \ref{tab:Avg_CC} lists only the  average Pearson correlation coefficients  for C$_\alpha$ B-factor predictions, which are obtained in the same manner as other heavy atoms. These results can be compared with those of 
 the parameter-free flexibility-rigidity index (pfFRI), Gaussian network model (GNM) and normal mode analysis (NMA) which, however, were obtained via the least squares fitting of each protein.}

\rd{Results for all heavy atom B-factor predictions for small-, medium-, and large-sized protein data subsets \cite{JKPark:2013} are given  in Tables \ref{tab:Small}, \ref{tab:Medium}, and \ref{tab:Large}.  Table \ref{tab:all} shows the results for all heavy atom B-factor predictions of each protein in the Superset. The average Pearson correlation coefficient for the data subsets is provided in Table \ref{tab:Avg_CC}. All methods perform similarly for the different protein data subsets with the convolutional neural network method performing the best on the Superset for both all heavy atom and C$_\alpha$ only B-factor predictions.}

{\small
\begin{table}[H]
\centering
\begin{tabular}{ccccccc}
\caption{Average Pearson correlation coefficients (PCC) both of all heavy atom and C$_\alpha$ only B-factor predictions for small-, medium-, and large-sized protein sets along with the entire superset of the 364 protein dataset. Predictions of random forest (RF), gradient boosted tree (GBT), and convolutional neural network (CNN) are obtained by leave-one-protein-out (blind), while predictions of parameter-free flexibility-rigidity index (pfFRI), Gaussian network model (GNM) and normal mode analysis (NMA) were obtained via the least squares fitting of individual proteins. All machine learning models use all heavy atom information for training.
\label{tab:Avg_CC}} \\
\hline\hline \\
\multicolumn{7}{c}{Prediction Of Only C$_{\alpha}$}\\ \hline	
\ \\
Protein Set	&	RF	&	GBT	&	CNN	&	 pfFRI \cite{Opron:2014} 	&	 GNM \cite{Opron:2014}	&	 NMA \cite{Opron:2014}\\ \hline 
							Small	&	0.25	&	0.39	&	0.53	&	0.60	&	0.54	&	 0.48\\
Medium	&	0.47	&	0.59	&	0.55	&	0.61	&	0.55	&	 0.48\\
Large	&	0.50	&	0.57	&	0.62	&	0.59	&	0.53	&	 0.49\\
Superset	&	0.49	&	0.57	&	0.66	&	0.63	&	0.57	&	 NA\\ \hline \hline
\multicolumn{7}{c}{Prediction Of All Heavy Atom}\\ \hline		
Protein Set	&	RF	&	GBT	&	CNN	&	 pfFRI \cite{Opron:2014} 	&	 GNM \cite{Opron:2014}	&	 NMA \cite{Opron:2014}\\ \hline 
							Small	&	0.44	&	0.49	&	0.56	&	 NA 	&	 NA 	&	 NA \\
Medium	&	0.59	&	0.64	&	0.62	&	 NA 	&	 NA 	&	 NA \\
Large	&	0.62	&	0.65	&	0.68	&	 NA 	&	 NA 	&	 NA \\
Superset	&	0.59	&	0.63	&	0.69	&	NA	&	NA	&	 NA\\ \hline \hline
\end{tabular}
\end{table}
}

\rd{Our blind prediction result using the convolutional neural network is notable because it improves upon the best result in the previous work for single protein parameters-free FRI (pfFRI) linear fitting of 0.63 \cite{Opron:2014 }. It is noted that blind predictions are much more difficult than linear fittings.  The result for single protein GNM linear fittings of the same data set is 0.57  \cite{Opron:2014 }. As reported in Table \ref{tab:all}, for each protein, no method outperforms any other method over the entire data set. In terms of the average Pearson correlation coefficient for all heavy atom B-factor prediction, the convolutional neural network method outperforms the boosted gradient and random forest by 10\% and 17\% respectively.}

\begin{table}[H]
\centering
\begin{tabular}{ccccc}

\caption{Pearson correlation coefficients for cross protein heavy atom blind B-factor prediction obtained by random forest (RF), boosted gradient (GBT), and convolutional neural network (CNN) for the small-sized protein set. Results reported use heavy atoms in both training and prediction. \label{tab:Small}} \\
\hline\hline \\
PDB ID	&	N	&	RF	&	GBT	&	CNN	\\ \hline 
1AIE	&	235	&	0.62	&	0.53	&	0.60	\\
1AKG	&	108	&	0.41	&	0.51	&	0.70	\\
1BX7	&	345	&	0.55	&	0.67	&	0.63	\\
1ETL	&	76	&	0.27	&	0.03	&	0.48	\\
1ETM	&	80	&	0.46	&	0.13	&	0.48	\\
1ETN	&	77	&	0.33	&	0.25	&	0.20	\\
1FF4	&	477	&	0.55	&	0.59	&	0.76	\\
1GK7	&	321	&	0.53	&	0.73	&	0.72	\\
1GVD	&	401	&	0.66	&	0.69	&	0.71	\\
1HJE	&	73	&	-0.07	&	0.46	&	0.37	\\
1KYC	&	138	&	0.43	&	0.30	&	0.32	\\
1NOT	&	96	&	-0.18	&	0.81	&	0.63	\\
1O06	&	142	&	0.51	&	0.64	&	0.65	\\
1P9I	&	203	&	0.73	&	0.77	&	0.77	\\
1PEF	&	153	&	0.60	&	0.64	&	0.76	\\
1PEN	&	109	&	0.34	&	0.24	&	0.21	\\
1Q9B	&	303	&	0.41	&	0.67	&	0.75	\\
1RJU	&	257	&	0.71	&	0.75	&	0.73	\\
1U06	&	432	&	0.55	&	0.68	&	0.61	\\
1UOY	&	452	&	0.55	&	0.56	&	0.55	\\
1USE	&	290	&	0.25	&	0.50	&	0.68	\\
1VRZ	&	66	&	0.38	&	-0.17	&	0.09	\\
1XY2	&	62	&	0.16	&	0.27	&	0.55	\\
1YJO	&	55	&	0.36	&	0.12	&	0.02	\\
1YZM	&	361	&	0.51	&	0.60	&	0.56	\\
2DSX	&	386	&	0.36	&	0.44	&	0.56	\\
2JKU	&	229	&	0.57	&	0.63	&	0.35	\\
2NLS	&	269	&	0.45	&	0.49	&	0.70	\\
2OL9	&	51	&	0.65	&	0.51	&	0.84	\\
6RXN	&	345	&	0.56	&	0.71	&	0.82	\\ \hline\hline

\end{tabular}
\end{table}

\begin{table}[H]
\centering
\begin{tabular}{ccccc}
\caption{Pearson correlation coefficients for cross protein heavy atom blind B-factor prediction obtained by random forest (RF), boosted gradient (GBT), and convolutional neural network (CNN) for the medium-sized protein set. Results reported use heavy atoms in both training and prediction.\label{tab:Medium}} \\
\hline\hline \\
PDB ID	&	N	&	RF	&	GBT	&	CNN	\\ \hline 
1ABA	&	728	&	0.74	&	0.77	&	0.73	\\
1CYO	&	697	&	0.66	&	0.68	&	0.76	\\
1FK5	&	626	&	0.62	&	0.71	&	0.63	\\
1GXU	&	694	&	0.65	&	0.67	&	0.66	\\
1I71	&	683	&	0.57	&	0.62	&	0.66	\\
1LR7	&	522	&	0.53	&	0.70	&	0.71	\\
1N7E	&	700	&	0.62	&	0.65	&	0.71	\\
1NNX	&	674	&	0.69	&	0.73	&	0.53	\\
1NOA	&	778	&	0.52	&	0.57	&	0.57	\\
1OPD	&	642	&	0.55	&	0.60	&	0.62	\\
1QAU	&	812	&	0.57	&	0.58	&	0.57	\\
1R7J	&	729	&	0.71	&	0.70	&	0.65	\\
1UHA	&	623	&	0.74	&	0.80	&	0.75	\\
1ULR	&	677	&	0.69	&	0.71	&	0.68	\\
1USM	&	631	&	0.59	&	0.78	&	0.67	\\
1V05	&	17	&	-0.20	&	0.02	&	0.60	\\
1W2L	&	746	&	0.62	&	0.68	&	0.69	\\
1X3O	&	622	&	0.53	&	0.52	&	0.63	\\
1Z21	&	771	&	0.63	&	0.66	&	0.63	\\
1ZVA	&	551	&	0.59	&	0.56	&	0.58	\\
2BF9	&	287	&	0.39	&	0.52	&	0.70	\\
2BRF	&	735	&	0.76	&	0.78	&	0.86	\\
2CE0	&	714	&	0.62	&	0.65	&	0.90	\\
2E3H	&	589	&	0.70	&	0.73	&	0.38	\\
2EAQ	&	705	&	0.63	&	0.61	&	0.58	\\
2EHS	&	590	&	0.55	&	0.71	&	0.38	\\
2FQ3	&	721	&	0.67	&	0.75	&	0.76	\\
2IP6	&	702	&	0.62	&	0.67	&	0.64	\\
2MCM	&	735	&	0.71	&	0.73	&	0.60	\\
2NUH	&	806	&	0.64	&	0.72	&	0.19	\\
2PKT	&	666	&	0.06	&	0.17	&	0.76	\\
2PLT	&	719	&	0.62	&	0.67	&	0.70	\\
2QJL	&	734	&	0.61	&	0.60	&	0.42	\\
2RB8	&	723	&	0.61	&	0.64	&	0.42	\\
3BZQ	&	742	&	0.60	&	0.61	&	0.43	\\
5CYT	&	800	&	0.68	&	0.70	&	0.74	\\ \hline\hline
\end{tabular}
\end{table}

\begin{table}[H]
\centering
\caption{Pearson correlation coefficients for cross protein heavy atom blind B-factor prediction obtained by random forest (RF), boosted gradient (GBT), and convolutional neural network (CNN) for the large-sized protein set. Results reported use heavy atoms in both training and prediction.\label{tab:Large}} 
\begin{tabular}{ccccc}
\hline\hline \\
PDB ID	&	N	&	RF	&	GBT	&	CNN	\\ \hline
1AHO	&	482	&	0.62	&	0.71	&	0.76	\\
1ATG	&	1689&	0.61	&	0.66	&	0.63	\\
1BYI	&	1540&	0.59	&	0.63	&	0.59	\\
1CCR	&	837	&	0.70	&	0.67	&	0.66	\\
1E5K	&	1423&	0.70	&	0.73	&	0.74	\\
1EW4	&	863	&	0.70	&	0.71	&	0.61	\\
1IFR	&	878	&	0.72	&	0.74	&	0.73	\\
1NLS	&	1746&	0.61	&	0.64	&	0.56	\\
1O08	&	1722&	0.51	&	0.58	&	0.55	\\
1PMY	&	937	&	0.64	&	0.65	&	0.67	\\
1PZ4	&	874	&	0.73	&	0.73	&	0.74	\\
1QTO	&	934	&	0.61	&	0.55	&	0.63	\\
1RRO	&	846	&	0.56	&	0.52	&	0.54	\\
1UKU	&	873	&	0.74	&	0.75	&	0.70	\\
1V70	&	784	&	0.70	&	0.67	&	0.62	\\
1WBE	&	1542&	0.59	&	0.61	&	0.63	\\
1WHI	&	937	&	0.74	&	0.77	&	0.71	\\
1WPA	&	906	&	0.64	&	0.66	&	0.74	\\
2AGK	&	1867&	0.61	&	0.68	&	0.44	\\
2C71	&	1446&	0.59	&	0.61	&	0.83	\\
2CG7	&	536	&	0.47	&	0.54	&	0.79	\\
2CWS	&	1624&	0.63	&	0.60	&	0.78	\\
2HQK	&	1582&	0.76	&	0.76	&	0.90	\\
2HYK	&	1832&	0.60	&	0.65	&	0.85	\\
2I24	&	872	&	0.52	&	0.52	&	0.91	\\
2IMF	&	1564&	0.62	&	0.62	&	0.47	\\
2PPN	&	701	&	0.50	&	0.68	&	0.83	\\
2R16	&	1262&	0.52	&	0.53	&	0.50	\\
2V9V	&	986	&	0.64	&	0.61	&	0.63	\\
2VIM	&	781	&	0.62	&	0.61	&	0.75	\\
2VPA	&	1524&	0.63	&	0.68	&	0.61	\\
2VYO	&	1589&	0.53	&	0.65	&	0.61	\\
3SEB	&	1948&	0.61	&	0.71	&	0.57	\\
3VUB	&	787	&	0.64	&	0.70	&	0.78	\\ \hline\hline

\end{tabular}
\end{table}

{\small 
\begin{longtable}[H]{llccc|llccc}
\caption{Pearson correlation coefficients for cross protein heavy atom blind B-factor prediction obtained by random forest (RF), boosted gradient (GBT), and convolutional neural network (CNN) for the Superset. Results reported use heavy atoms in both training and prediction.\\ } \label{tab:365} \\
\hline\hline \multicolumn{1}{c}{\textbf{PDB}} & \multicolumn{1}{c}{\textbf{N}} & \multicolumn{1}{c}{\textbf{RF}} & \multicolumn{1}{c}{\textbf{GBT}} & \multicolumn{1}{c|}{\textbf{CNN}} &
\multicolumn{1}{c}{\textbf{PDB}} & \multicolumn{1}{c}{\textbf{N}} & \multicolumn{1}{c}{\textbf{RF}} & \multicolumn{1}{c}{\textbf{GBT}} & \multicolumn{1}{c}{\textbf{CNN}} 
\\ \hline 
\endfirsthead

\multicolumn{10}{c}%
{{\bfseries \tablename\ \thetable{} -- continued from previous page}} \\ \hline
\hline \multicolumn{1}{c}{\textbf{PDB}} & \multicolumn{1}{c}{\textbf{N}} & \multicolumn{1}{c}{\textbf{RF}} & \multicolumn{1}{c}{\textbf{GBT}} & \multicolumn{1}{c|}{\textbf{CNN}} &
\multicolumn{1}{c}{\textbf{PDB}} & \multicolumn{1}{c}{\textbf{N}} & \multicolumn{1}{c}{\textbf{RF}} & \multicolumn{1}{c}{\textbf{GBT}} & \multicolumn{1}{c}{\textbf{CNN}} 
\\ \hline 
\endhead

\hline \multicolumn{3}{|r|}{{Continued on next page}} \\ \hline
\endfoot

\hline \hline
\endlastfoot

 \hline \\
1ABA&	728	&	0.74	&	0.77	&	0.73	&	2X5Y	&	1352	&	0.75	&	0.79	&	0.72	\\
1AHO&	482	&	0.62	&	0.71	&	0.76	&	2X9Z	&	1956	&	0.71	&	0.72	&	0.76	\\
1AIE&	235	&	0.62	&	0.53	&	0.60	&	2XHF	&	2432	&	0.65	&	0.71	&	0.70	\\
1AKG&	108	&	0.41	&	0.51	&	0.70	&	2Y0T	&	757	&	0.59	&	0.75&	0.73	\\
1ATG&	1689	&	0.61	&	0.66	&	0.63	&	2Y72	&	1171	&	0.73	&	0.80	&	0.75	\\
1BGF&	1018	&	0.58	&	0.63	&	0.63	&	2Y7L	&	2398	&	0.81	&	0.82	&	0.62	\\
1BX7&	345	&	0.55	&	0.67	&	0.63	&	2Y9F	&	1212	&	0.72	&	0.77	&	0.64	\\
1BYI&	1540	&	0.59	&	0.63	&	0.59	&	2YLB	&	3065	&	0.60	&	0.69	&	0.63	\\
1CCR&	837	&	0.70	&	0.67	&	0.66	&	2YNY	&	2364	&	0.67	&	0.71	&	0.68	\\
1CYO&	697	&	0.66	&	0.68	&	0.76	&	2ZCM	&	2959	&	0.41	&	0.45	&	0.44	\\
1DF4&	463	&	0.79	&	0.75	&	0.64	&	2ZU1	&	2794	&	0.59	&	0.73	&	0.17	\\
1E5K&	1423	&	0.70	&	0.73	&	0.74	&	3A0M	&	823	&	0.65	&	0.47	&	0.74	\\
1ES5&	1912	&	0.63	&	0.68	&	0.66	&	3A7L	&	963	&	0.66	&	0.75	&	0.81	\\
1ETL&	76	&	0.27	&	0.03	&	0.48	&	3AMC	&	5174	&	0.72	&	0.75	&	0.62	\\
1ETM&	80	&	0.46	&	0.13	&	0.48	&	3AUB	&	782	&	0.63	&	0.62	&	0.74	\\
1ETN&	77	&	0.33	&	0.25	&	0.20	&	3B5O	&	1510	&	0.53	&	0.55	&	0.65	\\
1EW4	&	863	&	0.70	&	0.71	&	0.61	&	3BA1	&	2391	&	0.65	&	0.64	&	0.44	\\
1F8R	&	15291	&	0.64	&	0.64	&	0.83	&	3BED	&	1570	&	0.73	&	0.73	&	0.70	\\
1FF4	&	477	&	0.55	&	0.59	&	0.76	&	3BQX	&	1028	&	0.52	&	0.59	&	0.85	\\
1FK5	&	626	&	0.62	&	0.71	&	0.63	&	3BZQ	&	742	&	0.60	&	0.61	&	0.43	\\
1GCO	&	7888	&	0.64	&	0.61	&	0.71	&	3BZZ	&	773	&	0.45	&	0.45	&	0.77	\\
1GK7	&	321	&	0.53	&	0.73	&	0.72	&	3DRF	&	4101	&	0.67	&	0.66	&	0.81	\\
1GVD	&	401	&	0.66	&	0.69	&	0.71	&	3DWV	&	2363	&	0.60	&	0.67	&	0.87	\\
1GXU	&	694	&	0.65	&	0.67	&	0.66	&	3E5T	&	1543	&	0.71	&	0.72	&	0.75	\\
1H6V	&	22514	&	0.39	&	0.40	&	0.58	&	3E7R	&	295	&	0.60	&	0.60	&	0.81	\\
1HJE	&	73	&	-0.07	&	0.46	&	0.37	&	3EUR	&	1059	&	0.47	&	0.50	&	0.82	\\
1I71	&	683	&	0.57	&	0.62	&	0.66	&	3F2Z	&	1160	&	0.78	&	0.78	&	0.88	\\
1IDP	&	3661	&	0.69	&	0.74	&	0.83	&	3F7E	&	1912	&	0.61	&	0.67	&	0.69	\\
1IFR	&	878	&	0.72	&	0.74	&	0.73	&	3FCN	&	1039	&	0.68	&	0.71	&	0.73	\\
1K8U	&	686	&	0.65	&	0.68	&	0.74	&	3FE7	&	710	&	0.62	&	0.71	&	0.83	\\
1KMM	&	11632	&	0.65	&	0.70	&	0.87	&	3FKE	&	1938	&	0.57	&	0.56	&	0.76	\\
1KNG	&	1016	&	0.61	&	0.56	&	0.55	&	3FMY	&	470	&	0.73	&	0.75	&	0.84	\\
1KR4	&	906	&	0.73	&	0.76	&	0.72	&	3FOD	&	328	&	0.30	&	0.45	&	0.78	\\
1KYC	&	138	&	0.43	&	0.30	&	0.32	&	3FSO	&	197	&	0.71	&	0.73	&	0.85	\\
1LR7	&	522	&	0.53	&	0.70	&	0.71	&	3FTD	&	1795	&	0.75	&	0.75	&	0.69	\\
1MF7	&	1551	&	0.68	&	0.68	&	0.70	&	3G1S	&	3196	&	0.74	&	0.76	&	0.72	\\
1N7E	&	700	&	0.62	&	0.65	&	0.71	&	3GBW	&	1275	&	0.75	&	0.76	&	0.68	\\
1NKD	&	426	&	0.56	&	0.59	&	0.63	&	3GHJ	&	808	&	0.66	&	0.71	&	0.44	\\
1NLS	&	1746	&	0.61	&	0.64	&	0.56	&	3HFO	&	1432	&	0.65	&	0.72	&	0.70	\\
1NNX	&	674	&	0.69	&	0.73	&	0.53	&	3HHP	&	8495	&	0.71	&	0.74	&	0.62	\\
1NOA	&	778	&	0.52	&	0.57	&	0.57	&	3HNY	&	1351	&	0.73	&	0.73	&	0.58	\\
1NOT	&	96	&	-0.18	&	0.81	&	0.63	&	3HP4	&	1322	&	0.61	&	0.63	&	0.65	\\
1O06	&	142	&	0.51	&	0.64	&	0.65	&	3HWU	&	934	&	0.51	&	0.69	&	0.51	\\
1O08	&	1722	&	0.51	&	0.58	&	0.55	&	3HYD	&	52	&	-0.05	&	0.28	&	0.60	\\
1OPD	&	642	&	0.55	&	0.60	&	0.62	&	3HZ8	&	1459	&	0.51	&	0.54	&	0.76	\\
1P9I	&	203	&	0.73	&	0.77	&	0.77	&	3I2V	&	929	&	0.50	&	0.54	&	0.81	\\
1PEF	&	153	&	0.60	&	0.64	&	0.76	&	3I2Z	&	1039	&	0.63	&	0.64	&	0.75	\\
1PEN	&	109	&	0.34	&	0.24	&	0.21	&	3I4O	&	969	&	0.66	&	0.64	&	0.87	\\
1PMY	&	937	&	0.64	&	0.65	&	0.67	&	3I7M	&	928	&	0.56	&	0.60	&	0.87	\\
1PZ4	&	874	&	0.73	&	0.73	&	0.74	&	3IHS	&	1120	&	0.66	&	0.65	&	0.81	\\
1Q9B	&	303	&	0.41	&	0.67	&	0.75	&	3IVV	&	1097	&	0.72	&	0.81	&	0.85	\\
1QAU	&	812	&	0.57	&	0.58	&	0.57	&	3K6Y	&	1617	&	0.62	&	0.65	&	0.90	\\
1QKI	&	31154	&	0.44	&	0.27	&	0.84	&	3KBE	&	829	&	0.75	&	0.76	&	0.86	\\
1QTO	&	934	&	0.61	&	0.55	&	0.63	&	3KGK	&	1492	&	0.75	&	0.78	&	0.87	\\
1R29	&	971	&	0.61	&	0.73	&	0.72	&	3KZD	&	605	&	0.64	&	0.70	&	0.74	\\
1R7J	&	729	&	0.71	&	0.70	&	0.65	&	3L41	&	1735	&	0.73	&	0.76	&	0.88	\\
1RJU	&	257	&	0.71	&	0.75	&	0.73	&	3LAA	&	1112	&	0.54	&	0.46	&	0.89	\\
1RRO	&	846	&	0.56	&	0.52	&	0.54	&	3LAX	&	753	&	0.69	&	0.71	&	0.89	\\
1SAU	&	830	&	0.62	&	0.68	&	0.60	&	3LG3	&	6061	&	0.57	&	0.59	&	0.91	\\
1TGR	&	749	&	0.61	&	0.65	&	0.67	&	3LJI	&	1946	&	0.46	&	0.54	&	0.50	\\
1TZV	&	1051	&	0.75	&	0.77	&	0.75	&	3M3P	&	1858	&	0.57	&	0.62	&	0.68	\\
1U06	&	432	&	0.55	&	0.68	&	0.61	&	3M8J	&	1396	&	0.78	&	0.77	&	0.68	\\
1U7I	&	1988	&	0.73	&	0.75	&	0.77	&	3M9J	&	1329	&	0.66	&	0.74	&	0.50	\\
1U9C	&	1712	&	0.61	&	0.64	&	0.58	&	3M9Q	&	1359	&	0.52	&	0.53	&	0.48	\\
1UHA	&	623	&	0.74	&	0.80	&	0.75	&	3MAB	&	1311	&	0.63	&	0.65	&	0.59	\\
1UKU	&	873	&	0.74	&	0.75	&	0.70	&	3MD4	&	81	&	0.36	&	0.61	&	0.79	\\
1ULR	&	677	&	0.69	&	0.71	&	0.68	&	3MEA	&	1236	&	0.58	&	0.64	&	0.93	\\
1UOY	&	452	&	0.55	&	0.56	&	0.55	&	3MGN	&	2236	&	0.15	&	0.03	&	0.82	\\
1USE	&	290	&	0.25	&	0.50	&	0.68	&	3MRE	&	2598	&	0.57	&	0.56	&	0.84	\\
1USM	&	631	&	0.59	&	0.78	&	0.67	&	3N11	&	2501	&	0.52	&	0.57	&	0.85	\\
1UTG	&	548	&	0.58	&	0.55	&	0.62	&	3NE0	&	1551	&	0.68	&	0.69	&	0.85	\\
1V05	&	17	&	-0.20	&	0.02	&	0.60	&	3NGG	&	702	&	0.63	&	0.75	&	0.83	\\
1V70	&	784	&	0.70	&	0.67	&	0.62	&	3NPV	&	3655	&	0.70	&	0.75	&	0.84	\\
1VRZ	&	66	&	0.38	&	-0.17	&	0.09	&	3NVG	&	50	&	-0.08	&	0.08	&	0.88	\\
1W2L	&	746	&	0.62	&	0.68	&	0.69	&	3NZL	&	567	&	0.59	&	0.65	&	0.63	\\
1WBE	&	1542	&	0.59	&	0.61	&	0.63	&	3O0P	&	1452	&	0.55	&	0.65	&	0.63	\\
1WHI	&	937	&	0.74	&	0.77	&	0.71	&	3O5P	&	819	&	0.53	&	0.63	&	0.70	\\
1WLY	&	2430	&	0.65	&	0.71	&	0.68	&	3OBQ	&	1195	&	0.61	&	0.61	&	0.84	\\
1WPA	&	906	&	0.64	&	0.66	&	0.74	&	3OQY	&	1772	&	0.57	&	0.62	&	0.76	\\
1X3O	&	622	&	0.53	&	0.52	&	0.63	&	3P6J	&	857	&	0.57	&	0.70	&	0.88	\\
1XY1	&	124	&	0.58	&	0.19	&	0.47	&	3PD7	&	1354	&	0.70	&	0.72	&	0.85	\\
1XY2	&	62	&	0.16	&	0.27	&	0.55	&	3PES	&	1240	&	0.72	&	0.73	&	0.84	\\
1Y6X	&	669	&	0.44	&	0.53	&	0.46	&	3PID	&	3078	&	0.49	&	0.56	&	0.86	\\
1YJO	&	55	&	0.36	&	0.12	&	0.02	&	3PIW	&	1223	&	0.72	&	0.75	&	0.87	\\
1YZM	&	361	&	0.51	&	0.60	&	0.56	&	3PKV	&	1688	&	0.66	&	0.68	&	0.81	\\
1Z21	&	771	&	0.63	&	0.66	&	0.63	&	3PSM	&	729	&	0.62	&	0.68	&	0.80	\\
1ZCE	&	1100	&	0.77	&	0.81	&	0.73	&	3PTL	&	2101	&	0.61	&	0.62	&	0.72	\\
1ZVA	&	551	&	0.59	&	0.56	&	0.58	&	3PVE	&	2656	&	0.56	&	0.61	&	0.46	\\
2A50	&	3493	&	0.64	&	0.48	&	0.68	&	3PZ9	&	2913	&	0.63	&	0.76	&	0.60	\\
2AGK	&	1867	&	0.61	&	0.68	&	0.44	&	3PZZ	&	76	&	0.47	&	0.25	&	0.85	\\
2AH1	&	7215	&	0.65	&	0.57	&	0.67	&	3Q2X	&	43	&	0.29	&	0.59	&	0.76	\\
2B0A	&	1454	&	0.66	&	0.68	&	0.72	&	3Q6L	&	1022	&	0.71	&	0.67	&	0.75	\\
2BCM	&	3002	&	0.51	&	0.62	&	0.85	&	3QDS	&	2234	&	0.71	&	0.72	&	0.71	\\
2BF9	&	287	&	0.39	&	0.52	&	0.70	&	3QPA	&	1348	&	0.43	&	0.44	&	0.71	\\
2BRF	&	735	&	0.76	&	0.78	&	0.86	&	3R6D	&	1550	&	0.31	&	0.69	&	0.59	\\
2C71	&	1446	&	0.59	&	0.61	&	0.83	&	3R87	&	1007	&	0.39	&	0.51	&	0.53	\\
2CE0	&	714	&	0.62	&	0.65	&	0.90	&	3RQ9	&	1174	&	0.32	&	0.47	&	0.66	\\
2CG7	&	536	&	0.47	&	0.54	&	0.79	&	3RY0	&	964	&	0.66	&	0.65	&	0.53	\\
2COV	&	4366	&	0.76	&	0.83	&	0.78	&	3RZY	&	985	&	0.69	&	0.69	&	0.64	\\
2CWS	&	1624	&	0.63	&	0.60	&	0.78	&	3S0A	&	884	&	0.55	&	0.61	&	0.61	\\
2D5W	&	9772	&	0.71	&	0.75	&	0.75	&	3SD2	&	527	&	0.38	&	0.52	&	0.71	\\
2DKO	&	1933	&	0.71	&	0.72	&	0.72	&	3SEB	&	1948	&	0.61	&	0.71	&	0.57	\\
2DPL	&	4454	&	0.49	&	0.53	&	0.73	&	3SED	&	933	&	0.70	&	0.71	&	0.72	\\
2DSX	&	386	&	0.36	&	0.44	&	0.56	&	3SO6	&	1119	&	0.69	&	0.75	&	0.01	\\
2E10	&	3416	&	0.50	&	0.64	&	0.61	&	3SR3	&	4891	&	0.69	&	0.69	&	0.45	\\
2E3H	&	589	&	0.70	&	0.73	&	0.38	&	3SUK	&	1761	&	0.62	&	0.65	&	0.59	\\
2EAQ	&	705	&	0.63	&	0.61	&	0.58	&	3SZH	&	5074	&	0.74	&	0.80	&	0.44	\\
2EHP	&	1875	&	0.75	&	0.74	&	0.74	&	3T0H	&	1627	&	0.78	&	0.81	&	0.65	\\
2EHS	&	590	&	0.55	&	0.71	&	0.38	&	3T3K	&	922	&	0.56	&	0.68	&	0.62	\\
2ERW	&	385	&	0.47	&	0.50	&	0.32	&	3T47	&	1116	&	0.54	&	0.62	&	0.74	\\
2ETX	&	3018	&	0.56	&	0.61	&	0.58	&	3TDN	&	2703	&	0.55	&	0.55	&	0.58	\\
2FB6	&	766	&	0.63	&	0.65	&	0.52	&	3TOW	&	1193	&	0.53	&	0.66	&	0.66	\\
2FG1	&	1021	&	0.55	&	0.65	&	0.68	&	3TUA	&	1510	&	0.63	&	0.66	&	0.70	\\
2FN9	&	4362	&	0.37	&	0.60	&	0.61	&	3TYS	&	556	&	0.67	&	0.68	&	0.71	\\
2FQ3	&	721	&	0.67	&	0.75	&	0.76	&	3U6G	&	1658	&	0.52	&	0.51	&	0.60	\\
2G69	&	744	&	0.60	&	0.61	&	0.87	&	3U97	&	524	&	0.57	&	0.66	&	0.27	\\
2G7O	&	537	&	0.52	&	0.63	&	0.89	&	3UCI	&	536	&	0.44	&	0.51	&	0.56	\\
2G7S	&	1258	&	0.60	&	0.60	&	0.81	&	3UR8	&	5033	&	0.63	&	0.66	&	0.83	\\
2GKG	&	706	&	0.63	&	0.60	&	0.70	&	3US6	&	1156	&	0.62	&	0.64	&	0.01	\\
2GOM	&	987	&	0.61	&	0.70	&	0.92	&	3V1A	&	319	&	0.36	&	0.36	&	0.76	\\
2GXG	&	1132	&	0.67	&	0.75	&	0.86	&	3V75	&	1974	&	0.63	&	0.65	&	0.83	\\
2GZQ	&	1402	&	0.59	&	0.60	&	0.90	&	3VN0	&	1469	&	0.69	&	0.76	&	0.76	\\
2HQK	&	1582	&	0.76	&	0.76	&	0.90	&	3VOR	&	1077	&	0.41	&	0.50	&	0.81	\\
2HYK	&	1832	&	0.60	&	0.65	&	0.85	&	3VUB	&	787	&	0.64	&	0.70	&	0.78	\\
2I24	&	872	&	0.52	&	0.52	&	0.91	&	3VVV	&	869	&	0.62	&	0.69	&	0.84	\\
2I49	&	3109	&	0.78	&	0.77	&	0.90	&	3VZ9	&	1366	&	0.70	&	0.72	&	0.66	\\
2IBL	&	815	&	0.46	&	0.53	&	0.88	&	3W4Q	&	5406	&	0.66	&	0.73	&	0.65	\\
2IGD	&	431	&	0.58	&	0.68	&	0.82	&	3ZBD	&	1718	&	0.54	&	0.54	&	0.78	\\
2IMF	&	1564	&	0.62	&	0.62	&	0.47	&	3ZIT	&	1192	&	0.51	&	0.54	&	0.71	\\
2IP6	&	702	&	0.62	&	0.67	&	0.64	&	3ZRX	&	1654	&	0.38	&	0.67	&	0.60	\\
2IVY	&	727	&	0.47	&	0.59	&	0.62	&	3ZSL	&	925	&	0.61	&	0.64	&	0.69	\\
2J32	&	1935	&	0.79	&	0.78	&	0.70	&	3ZZP	&	585	&	0.40	&	0.46	&	0.56	\\
2J9W	&	1626	&	0.66	&	0.68	&	0.73	&	3ZZY	&	1741	&	0.64	&	0.69	&	0.69	\\
2JKU	&	229	&	0.57	&	0.63	&	0.35	&	4A02	&	1281	&	0.62	&	0.65	&	0.75	\\
2JLI	&	708	&	0.58	&	0.54	&	0.73	&	4ACJ	&	1210	&	0.64	&	0.67	&	0.75	\\
2JLJ	&	889	&	0.66	&	0.70	&	0.68	&	4AE7	&	1458	&	0.64	&	0.74	&	0.61	\\
2MCM	&	735	&	0.71	&	0.73	&	0.60	&	4AM1	&	2605	&	0.64	&	0.67	&	0.56	\\
2NLS	&	269	&	0.45	&	0.49	&	0.70	&	4ANN	&	1180	&	0.53	&	0.60	&	0.72	\\
2NR7	&	1556	&	0.71	&	0.70	&	0.66	&	4AVR	&	1437	&	0.62	&	0.61	&	0.64	\\
2NUH	&	806	&	0.64	&	0.72	&	0.19	&	4AXY	&	317	&	0.45	&	0.64	&	0.75	\\
2O6X	&	2415	&	0.76	&	0.82	&	0.63	&	4B6G	&	4504	&	0.78	&	0.76	&	0.84	\\
2OA2	&	970	&	0.54	&	0.53	&	0.92	&	4B9G	&	2226	&	0.79	&	0.81	&	0.83	\\
2OHW	&	2074	&	0.55	&	0.62	&	0.81	&	4DD5	&	2618	&	0.63	&	0.66	&	0.87	\\
2OKT	&	2587	&	0.56	&	0.59	&	0.89	&	4DKN	&	3356	&	0.76	&	0.77	&	0.88	\\
2OL9	&	51	&	0.65	&	0.51	&	0.84	&	4DND	&	755	&	0.66	&	0.73	&	0.85	\\
2PKT	&	666	&	0.06	&	0.17	&	0.76	&	4DPZ	&	865	&	0.65	&	0.66	&	0.83	\\
2PLT	&	719	&	0.62	&	0.67	&	0.70	&	4DQ7	&	2526	&	0.58	&	0.69	&	0.78	\\
2PMR	&	590	&	0.63	&	0.66	&	0.63	&	4DT4	&	1163	&	0.71	&	0.73	&	0.73	\\
2POF	&	3418	&	0.58	&	0.66	&	0.85	&	4EK3	&	2147	&	0.70	&	0.72	&	0.73	\\
2PPN	&	701	&	0.50	&	0.68	&	0.83	&	4ERY	&	2357	&	0.70	&	0.74	&	0.83	\\
2PSF	&	4983	&	0.54	&	0.55	&	0.79	&	4ES1	&	737	&	0.63	&	0.64	&	0.81	\\
2PTH	&	1437	&	0.68	&	0.72	&	0.79	&	4EUG	&	1789	&	0.59	&	0.66	&	0.79	\\
2Q4N	&	9496	&	0.45	&	0.39	&	0.85	&	4F01	&	3374	&	0.55	&	0.54	&	0.77	\\
2Q52	&	26784	&	0.63	&	0.62	&	0.77	&	4F3J	&	1116	&	0.58	&	0.62	&	0.53	\\
2QJL	&	734	&	0.61	&	0.60	&	0.42	&	4FR9	&	956	&	0.61	&	0.64	&	0.62	\\
2R16	&	1262	&	0.52	&	0.53	&	0.50	&	4G14	&	39	&	0.28	&	0.50	&	0.55	\\
2R6Q	&	903	&	0.59	&	0.53	&	0.57	&	4G2E	&	1178	&	0.73	&	0.73	&	0.76	\\
2RB8	&	723	&	0.61	&	0.64	&	0.42	&	4G5X	&	4002	&	0.74	&	0.75	&	0.65	\\
2RE2	&	1559	&	0.66	&	0.66	&	0.54	&	4G6C	&	4814	&	0.47	&	0.60	&	0.61	\\
2RFR	&	1019	&	0.54	&	0.58	&	0.66	&	4G7X	&	1315	&	0.49	&	0.56	&	0.80	\\
2V9V	&	986	&	0.64	&	0.61	&	0.63	&	4GA2	&	873	&	0.51	&	0.55	&	0.55	\\
2VE8	&	3967	&	0.65	&	0.59	&	0.66	&	4GMQ	&	678	&	0.56	&	0.72	&	0.54	\\
2VH7	&	749	&	0.74	&	0.70	&	0.82	&	4GS3	&	737	&	0.56	&	0.60	&	0.56	\\
2VIM	&	781	&	0.62	&	0.61	&	0.75	&	4H4J	&	1470	&	0.69	&	0.80	&	0.70	\\
2VPA	&	1524	&	0.63	&	0.68	&	0.61	&	4H89	&	1127	&	0.55	&	0.61	&	0.62	\\
2VQ4	&	800	&	0.72	&	0.76	&	0.78	&	4HDE	&	1288	&	0.73	&	0.79	&	0.70	\\
2VY8	&	1058	&	0.71	&	0.74	&	0.63	&	4HJP	&	2112	&	0.65	&	0.70	&	0.76	\\
2VYO	&	1589	&	0.53	&	0.65	&	0.61	&	4HWM	&	799	&	0.50	&	0.57	&	0.81	\\
2W1V	&	4223	&	0.68	&	0.72	&	0.72	&	4IL7	&	527	&	0.35	&	0.43	&	0.74	\\
2W2A	&	2918	&	0.56	&	0.62	&	0.63	&	4J11	&	2658	&	0.47	&	0.58	&	0.94	\\
2W6A	&	826	&	0.66	&	0.76	&	0.69	&	4J5O	&	1406	&	0.64	&	0.63	&	0.91	\\
2WJ5	&	630	&	0.49	&	0.53	&	0.77	&	4J5Q	&	1062	&	0.73	&	0.75	&	0.87	\\
2WUJ	&	828	&	0.55	&	0.55	&	0.55	&	4J78	&	2443	&	0.71	&	0.75	&	0.86	\\
2WW7	&	915	&	0.35	&	0.43	&	0.61	&	4JG2	&	1294	&	0.70	&	0.73	&	0.88	\\
2WWE	&	54	&	0.23	&	0.22	&	0.12	&	4JVU	&	1615	&	0.69	&	0.68	&	0.89	\\
2X1Q	&	1852	&	0.58	&	0.53	&	0.77	&	4JYP	&	4063	&	0.70	&	0.78	&	0.93	\\
2X25	&	1289	&	0.65	&	0.68	&	0.80	&	4KEF	&	1002	&	0.65	&	0.62	&	0.68	\\
2X3M	&	1267	&	0.66	&	0.70	&	0.75	&	5CYT	&	800	&	0.68	&	0.70	&	0.74	\\
	&		&		&		&		&	6RXN	&	345	&	0.56	&	0.71	&	0.82	\\

\label{tab:all}																							
\end{longtable}}

Some low Pearson correlation coefficients results show a poor model prediction. However, in almost every protein where one model performs poorly, another model performs satisfactorily. \rd{When the maximum correlation coefficient for each protein is considered among the three methods the average all heavy atom correlation coefficient is increased to 0.73 and the average C$_\alpha$ only correlation coefficient is increased to 0.72.} This result is similar to that of the parameter-optimized FRI (opFRI) reported in our earlier work  \cite{Opron:2014}.

\subsection{Relative feature importance}

\begin{figure}[H]
\centering
\includegraphics[width=1\textwidth]{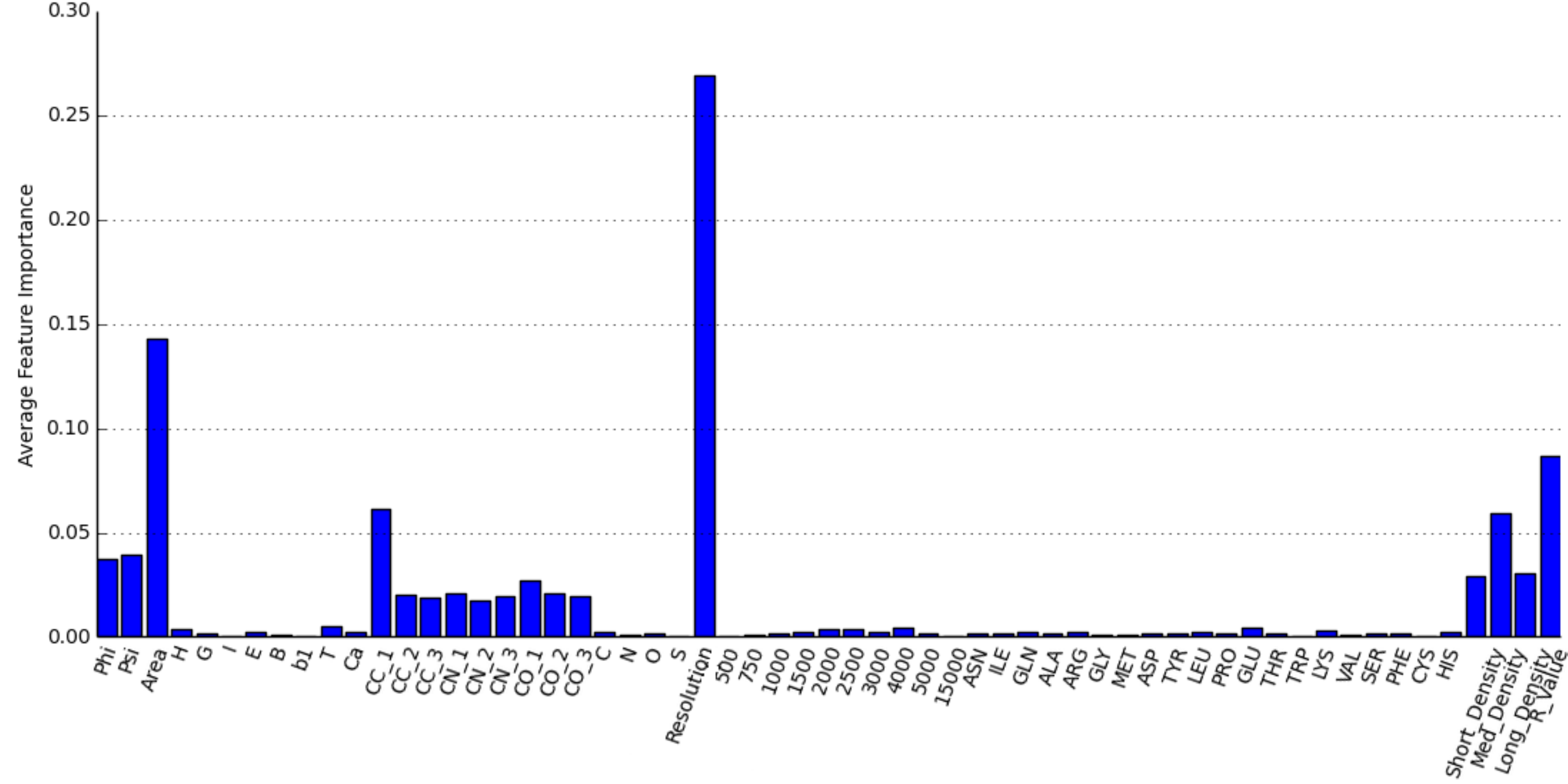}
\caption{Individual feature importance for the random forest model averaged over the data set. Reported feature selection includes the use heavy atoms in the model.}
\label{fig:RF_feat}
\end{figure}

Both random forest and boosted gradient methods have the ability to rank relative feature importance helping us understand significant features in the model. Figure \ref{fig:RF_feat} shows the individual feature importance for the random forest averaged over the dataset.

\begin{figure}[H]
\centering
\includegraphics[width=0.5\textwidth]{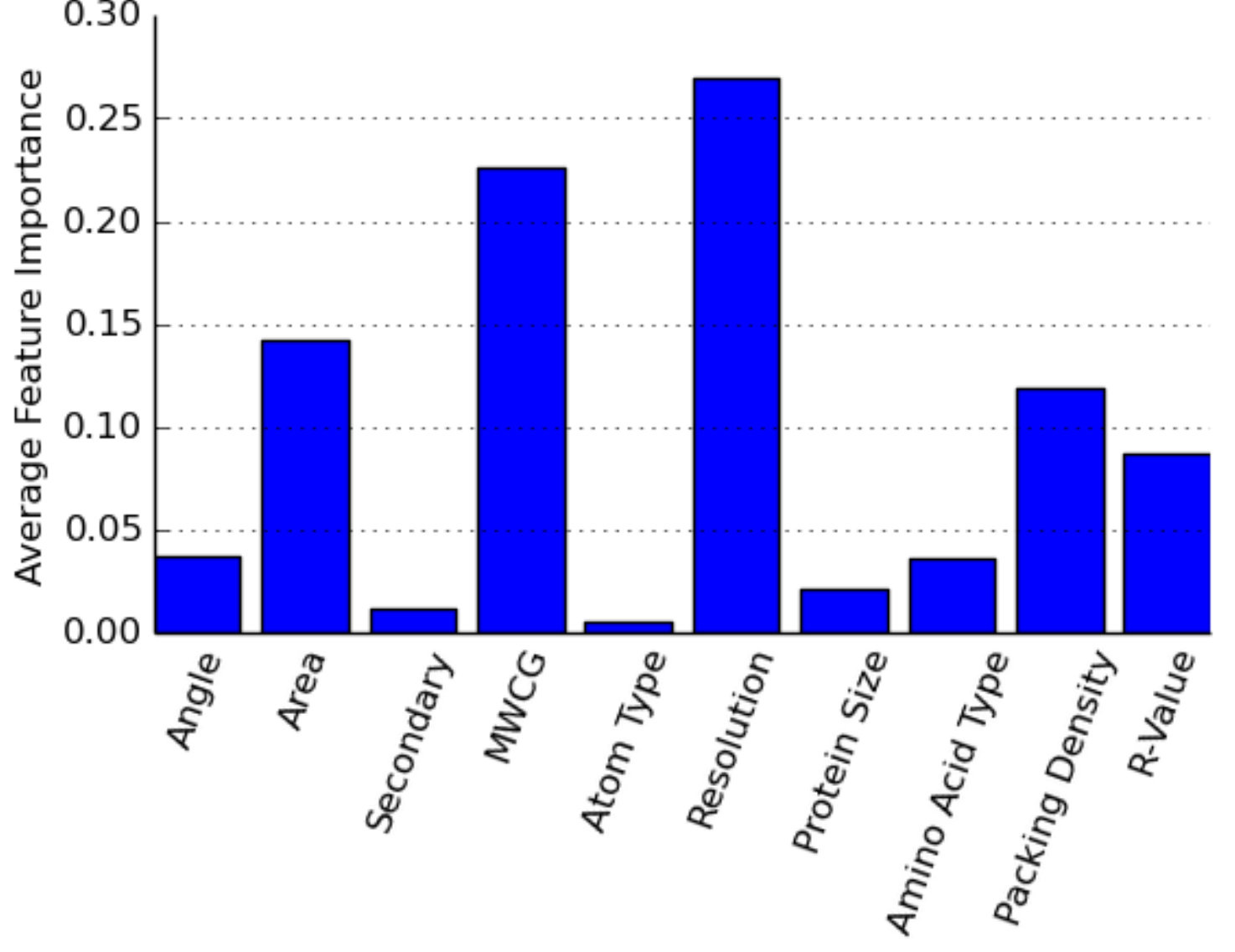}
\caption{Average feature importance for the random forest model with  the angle, secondary, MWCG, atom type, protein size, amino acid, and packing density features aggregated. Reported feature selection includes the use heavy atoms in the model.}
\label{fig:RF_feat_sum}
\end{figure}

We also include aggregated feature importance in Figure \ref{fig:RF_feat_sum}. In this figure, we sum the importance of the individual angle, secondary, MWCG, atom type, protein size, amino acid, and packing density features.

Figure \ref{fig:RF_feat} shows the most important MWCG feature is the carbon-carbon interaction. This MWCG feature uses a Lorentz radial basis function as with $\eta = 16$ and $\nu=3$ as detailed in Section \ref{sect:ML}.  The remaining eight MWCG features all rank similarly with the carbon-oxygen interaction ranked as the second most significant MWCG feature. This result validates that the model benefits from the multi-scale property of the MWCG feature, which uses three different kernels to capture interactions at various length scales. Since all MWCG have significance in the feature ranking it follows that the element specific property of the MWCG method is also a meaningful model feature.

Figure \ref{fig:RF_feat} shows that that the individual MWCG, amino acid type, and packing density feature have low relative importance, however, considering their aggregate importance as seen in Figure \ref{fig:RF_feat_sum}, we see that they contribute to the model. Figure  \ref{fig:RF_feat_sum} shows that the  medium density protein packing density feature was twice as important to the model as the short and long density features. The medium packing density may be capturing semi-local side chain interactions which are important in protein flexibility. The short packing density likely captures only adjacent backbone information while the long packing density is only adding weak atomic interaction information to the model. Protein resolution is the most significant relative feature followed by MWCG features and the STRIDE generated residue solvent accessible area feature. This also highlights the importance of the quality of X-ray crystal structures and difficulty in cross-protein B-factor prediction. Protein angles, secondary structures, and size play a less significant role in the model compared to the other features. Atom type has the lowest significance relative to the other features implemented in the model. Not surprisingly, we see that global features such as resolution and R-value are important components in the ensemble model. The global feature of protein size has a small role in the model.

Care must be taken to use feature ranking to understand feature importance. The feature ranking provided by these models is a relative ordering of features that the models find most important. So features with high correlation may be redundant giving one of them  a lower rank even though they may have significant prediction power. For example, R-value highly correlates with resolution so it is likely a meaningful feature. However,  the use of resolution reduces  the relative importance ranking of R-value in the model.

\subsection{Machine learning methods}\label{sect:Dis}

\rd{Among the three methods considered in this work, the convolutional neural network method outperforms the boosted gradient tree and random forest by 10\% and 17\%, respectively.}  As reported in Table \ref{tab:all}, no machine learning method outperforms any other method for each of all proteins. Results for all machine learning methods could undoubtedly be improved by refining features, exploring new features, and further tuning hyperparameters. 

In general, ensemble methods do not require as much parameter tuning as the CNN does.  The random forest is the simplest and most robust method. To balance cost, time, and quality only 500 trees were used for the random forest and 1000 trees were used for the boosted gradient method in this work. This may account for the increased performance of the boosted gradient tree method compared to the random forest. Ensemble methods are quite robust against overfitting so adding more features would likely improve their results \cite{ZXCang:2017a}.  The boosted gradient trees  use several hyperparameters so these methods could benefit by further tuning these hyperparameters. 

\rd{The additional data in the form of MWCG images used in the convolutional neural network likely explains the improved performance as compared to the ensemble methods. More refined images and other novel image types could further improve results.}

Using the dropout strategy, CNNs are also robust against overfitting.  Since there are a few hyperparameters in the CNN method,  it would likely benefit from more detailed parameter tuning. Additionally, a large dataset and more features would also improve the CNN performance. For example, including persistent homology \cite{ZXCang:2017b} and differential geometry features might lead to a better CNN prediction. 


\section{Conclusion}\label{sect:con}
Protein flexibility is known to strongly correlate with protein function and its prediction is important for our
understanding of protein dynamics and transport. Our quantitative understanding of protein 
flexibility and function is greatly impeded by their complexity and a large number of degrees of freedom. Many time-independent methods,  such as NMA \cite{Brooks:1983,Go:1983,Levitt:1985,Tasumi:1982}, ENM \cite{Tirion:1996}, GNM \cite{Bahar:1997,Bahar:1998,Brooks:1983harmonic}, and FRI \cite{KLXia:2013f, Opron:2014,Opron:2015a,Opron:2016a}, exist that dramatically simplify the protein structural complexity and are able to analyze protein B-factors, which reflect protein flexibility among other things. Based on the hypothesis that intrinsic physics lies in a low-dimensional space embedded in a high-dimensional data space, we introduced multiscale weighted colored graphs (MWCGs) to effectively reduce protein structural complexity and efficiently describe protein flexibility. However, none of the aforementioned methods is able to blindly predict the protein B-factors of an unknown protein. This work integrates advanced machine learning algorithms and two sets of features, i.e., global and local ones, to blindly predict protein flexibility and B-factors.

A few standard datasets involving more than 300 proteins (or more than 600,000 of B-factors) have
been utilized to test the proposed method. \rd{We use the leave-one-protein-out scheme to blindly predict
protein B-factors of both all heavy atoms and only C$_\alpha$ atoms.} Extensive numerical experiments demonstrate that the present blind prediction is more
accurate than the least squares fitting using GNM or NMA in terms of Pearson's correlation coefficients for the prediction of C$_\alpha$ B-factor.  \rd{ Further, we demonstrate the ability to effectively blindly predict B-factors of any heavy atoms in a given protein.}

Three standard machine learning algorithms, namely, the random forest, graduate boosted trees, and
convolutional neural networks are employed in the present study. \rd{ Among them, convolutional neural networks
do a better job in B-factor predictions.} A variety of different features were considered for these models
including local, semi-local, and global features. Local features, such as MWCGs, are
designed to capture structural properties associated with the intrinsic flexibility while global features, such as X-ray crystal resolution, are used to enable the cross-protein comparison and analysis. The proposed method is very efficient. 
However, there is still much room for novel and interesting features that can be implemented in future work. For
example, many algebraic topology tools have been found very useful for protein analysis \cite{KLXia:2014c, ZXCang:2017a, ZXCang:2017b} and will likely pair well with machine learning approaches for protein 
flexibility predictions.

This work is a first step using the recent advances in machine learning techniques to blindly predict
protein B-factors. To the authors' knowledge, this is the first work demonstrating this as a feasible and
robust prediction method. This work provides a clear evidence that machine algorithms are useful in
protein  flexibility analysis.  Results for all methods could undoubtedly be improved by a better mathematical description of intrinsic flexibility, larger datasets, and more advanced machine learning algorithms.

The proposed methods could be implemented in a variety of interesting applications related to protein flexibility and function. These include topics such as hinge detection, hot spot identification, allosteric site detection,  pose prediction, protein folding, and computer-aided drug design.

\section*{Acknowledgment} 
\quad This work was supported in part by NSF Grants DMS-1721024  and DMS-1761320, NIH grant 1R01GM126189-01A1.

\vspace{1cm}
\clearpage


%

\end{document}